\documentclass[runningheads]{llncs}
\usepackage[T1]{fontenc}
\usepackage{graphicx}
\usepackage{cite}
\begin{document}
\title{A CT-Based Airway Segmentation Using U$^2$-net Trained by the Dice Loss Function}
\titlerunning{Airway Segmentation Using U$^2$-net and Dice Loss}
\author{Kunpeng Wang \and
Yuexi Dong \and
Yunpu Zeng \and Zhichun Ye \and Yangzhe Wang}
\authorrunning{K. Wang et al.}
\institute{
Sichuan University - Pittsburgh Institute, Chengdu, China \\
\email{kunpeng.wang@scupi.cn, chelseadong@stu.scu.edu.cn,
zengyunpu@stu.scu.edu.cn,yezhichun@stu.scu.edu.cn, wangyanzhe@stu.scu.edu.cn} }

\maketitle              % typeset the header of the contribution
\begin{abstract}
Airway segmentation from chest computed tomography scans has played an essential role in the pulmonary disease diagnosis. The computer-assisted airway segmentation based on the U-net architecture is more efficient and  accurate compared to the manual segmentation. In this paper we employ the U$^2$-net  
trained by the Dice loss function to model the airway tree from the multi-site CT scans based on 299 training CT scans provided by the ATM'22. The derived saliency probability map from the training 
is applied to the validation data to extract the corresponding airway trees. The observation shows that the majority of the segmented airway trees behave well from the perspective of accuracy and connectivity. Refinements such as non-airway regions labeling and removing are applied to certain obtained airway tree models to display the largest component of the binary results.

\end{abstract}
\section{Introduction}
The airway tree is one of the crucial structures of the human respiratory system. The diagnosis and assessment of many pulmonary diseases, such as asthma, bronchiectasis, and emphysema, depend on the careful examination on the structure of the airway tree. %Based on thoracic X-Ray computed tomography (CT), the manual segmentation of the airway tree is time-consuming and inaccurate due to the complex 3D tubular structure companied with small branches. 
The airway tree segmentation on CT scans belongs to the fine-grained image segmentation. Manual segmentation is usually time-consuming, inaccurate, and depends on the clinical experience of physicians, due to which, computer aids are mostly used in clinical practice to help complete the pulmonary airway segmentation . 

The development of the convolutional neural networks permits the effective automatic segmentation of airway trees. The U-net architecture as well as different variants based on the original U-net have become the most up-to-date and widely  applied convolutional neural network in the filed of airway segmentation \cite{ref_article4,ref_article5,ref_conf2}.

The main challenge of the segmentation of the airway tree is the celebrated input imbalance problem \cite{ref_article3,ref_article5} since the airway tree normally occupies less pixel/voxel values in a CT scan. In order to further improve the segmentation performance in terms of the extraction of not only the most obvious features at the main trachea but also the small features at the bronchi, we employ the U$^2$-net, which is a two-level nested U-architecture proposed by Qin \emph{et al.} \cite{ref_article1}, to segment the complex structure of the airway tree.  Note in the original architecture the U$^2$-net is proposed to supervise  by the binary cross entropy function. However, due to the imbalance problem  in the extraction, we train the U$^2$-net with the Dice loss function \cite{ref_conf3,ref_conf4}, which has proven to outperform at the imbalance problems.    With the 299 annotated CT scans provided by ATM'22 Grand Challenge \cite{ref_article2,ref_article3,ref_conf1}, we employ the U$^2$-net supervised by the Dice loss function  to obtain the most accurate possible airway tree modeling based on the optimization of the Dice Similarity Coefficient (DSC). %reach Dice Similarity Coefficient (DSC) maximization and multi-scale feature detection.

\section{The Methodology}

\begin{figure}[htbp!]
\includegraphics[width=\textwidth]{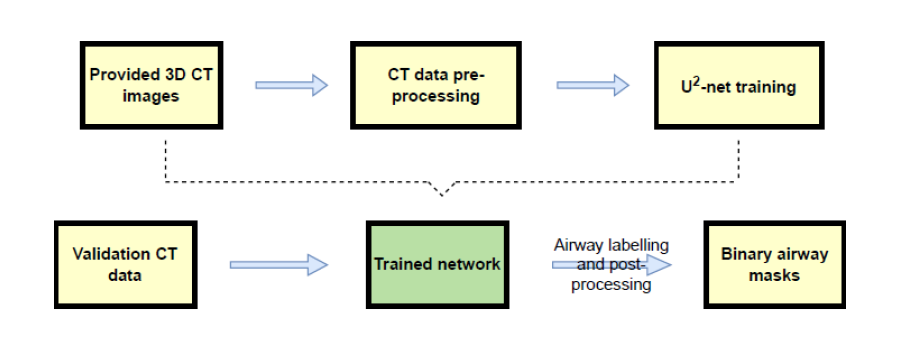}
\caption{Flowchart of the employed U$^2$-net segmentation.}
\label{fig3}
\end{figure}

In this section, we first introduce the preprocessing for the provided training data.  Next, we discuss the details about the employed U$^2$-net and the corresponding loss function. The final part of the section revolves around the generation of the binary masks for the validation data with the corresponding labeling and post-treatment. The airway segmentation treatment is illustrated in Fig. \ref{fig3}.

\subsection{Data Preprocessing}\label{sub1}
%This Challenge provides 299 CT scans with ground truth labels for training, 50 scans for validation, and 150 scans for testing. All 

The ATM'22  provides 499 three dimensional CT scans from multi-sites, of which 299 are given for training, 50 for validation and the rest 150 will be used by the challenge organizer for testing the submitted algorithms. The axial slices of the provided CT scans are of the size $512\times512$ in   length and width but with varying depth sizes. 

%First of all, we normalize the voxel values of each CT scan to accelerate the computation.

Given a chest training CT image, we first normalize the voxel values to prepare for faster computations in the following network training. The original distribution of the CT voxel intensity is around $[
-1024,3071]$.  We employ a CT-dependent normalization, that is, the intensity value is normalized according to the mean and standard deviation values of each slice. The normalized voxel intensity is rescaled to around $[0,2.5]$.  Note that the irrelevant dark parts with low intensities in a CT image, which normally enjoy the constant standard deviation value, are further reset to a different value, for example, $-9$ to distinguish from the rescaled array value $0$. 

In our training, we adopt the original size $512 \times 512$ of each slice with the GPU memory permitting. However, the depth sizes are doubled, that is, each input slice image is of $512 \times 512 \times 2$ to improve the performance accuracy of the employed U$^2$-net in the feature extraction. Although the treatment may demand more GPU memory, we have shown, with the comparison to the input images in the original depth size, that the double size treatment indeed significantly increases the training accuracy. 

%In order to capture features of airway trees at different levels of scale, we do not crop the axial slices and preserve the full size of $512\times512$. We then duplicate each slice and feed the $512\times512\times2$ slices as the input to the network for the purpose of better extraction of features. 

\subsection{Network Architecture}
In the training process we employ the U$^n$-net architecture \cite{ref_article1} to obtain the saliency probability function for the airway tree. The U$^n$-net is a n-level nested U-structure, where the exponent $n$ denotes the the level of the nested U-structures. We adopt $n=2$ in the application from the practical perspective by following \cite{ref_article1}. The U$^2$-net is built by two levels of U-structures. In details,  the top level U-structure consists of a  six-stage encoder, a five-stage decoder, and a saliency map fusion module and each stage among the encoders and decoders contains the newly proposed Residual U-block (abbreviated as RSU), which is the nested bottom level  downsampling upsampling encoder-decoder U-net. Each level of the downsampling/upsampling paths in the corresponding RSU consists of a $3\times3$ convolutional layer, a batch normalization layer, a rectified linear unit (ReLu) activation function, and a $2\times2$ max-pooling/upsampling layer. The nested RSU allows each block to use the multiscale features as residuals instead of the original features, which makes the whole architecture maintain high resolution feature with a reasonable requirement of the GPU memory and computing ability.

While the input data are processed by each RSU, the corresponding encoder and decoder with   a $3\times3$ convolution layer, followed by a sigmoid function in each layer of the upper level U-net will generate six side output saliency probability maps $S_{side}^{(6)}, S_{side}^{(5)}, S_{side}^{(4)}, S_{side}^{(3)}, S_{side}^{(2)}, S_{side}^{(1)}$,  by applying a $1\times1$ convolution layer and a sigmoid function to the concatenation  of which leads to a final saliency probability map $S_{fuse}$. In addition to the economical cost of GPU memory and computing capacity, the training of the U$^2$-net does not rely on any backbones due to the aforementioned nested design.

As mentioned in Subsection \ref{sub1}, the input is $512\times512\times2$ duplicated slices, hence, we change the channel number of the side probability output at each level to 3 to improve the performance of the architecture in extracting information and features at different levels and facilitate the recognition of the trachea as well as the small peripheral bronchi.

% Aiming for a better fusion of the multi-scale features, we modified the channel number of the side probability output at each level to be 3, which allows the ability to capture information and features at different levels and thus facilitates the recognition of the trachea as well as the small peripheral bronchi. The side output probability maps can help to preserve the detailed feature of bronchi as we go deeper in the network and downsample the feature map.

\subsection{The Loss Function}
The training of the U$^2$-net follows an end-to-end supervised manner, whose training loss is defined as the weighted sum of the loss of the side and final output saliency probability maps 
\begin{equation}
    L = \sum_{n=1}^{N} w_{side}^{(n)}l_{side}^{(n)}+w_{fuse}l_{fuse},
\end{equation}
where $l_{side}^{(n)}$ ($n=6$, since there are 6 side output in total) is the loss of the side output probability map, $l_{fuse}$ is the loss of the final output, and $w_{side}$ as well as ${w_{fuse}}$ are the associated weights of each loss. More specifically, we utilize the Dice loss function for each $l$ term (including both $l_{side}$ and $l_{fuse}$), 
\begin{equation}
    l = 1-\frac{2\sum\limits_{i=1}^{M} P_{G(i)}P_{S(i)} + \epsilon}{\sum\limits_{i=1}^{M} P_{G(i)}^2 + \sum\limits_{i=1}^{M} P_{S(i)}^2 +\epsilon},
\end{equation}
where $M$ denotes the total voxel index in a ground truth, $P_{G(i)}$ is the numerical array value of the ground truth and $P_{S(i)}$ represents the corresponding predicted value of the saliency probability map, $\epsilon$ is a coefficient used to avoid division by zero and we take $\epsilon=1$. The network is then trained via optimizing the overall loss $L$ by evaluating the network output with the given ground truth.

\subsection{Post-processing}\label{sub2}
\begin{figure}
\includegraphics[width=\textwidth]{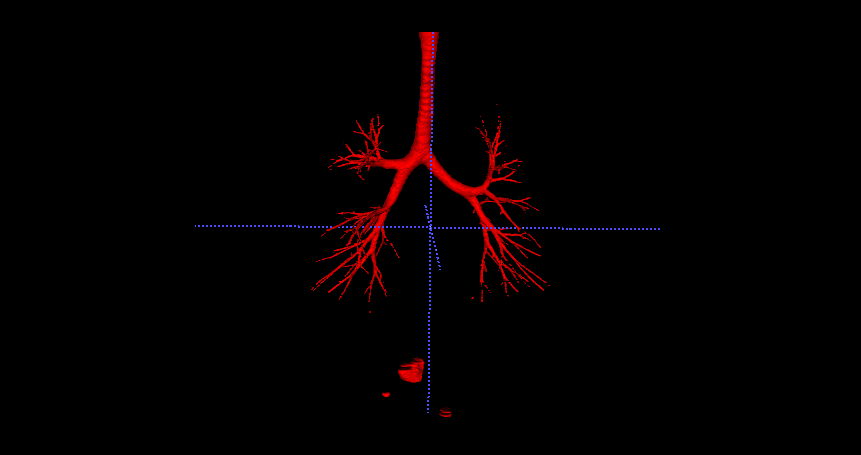}
\caption{Unrefined airway tree segmentation for the validation case ATM\textunderscore292\textunderscore0000.}
\label{fig4}
\end{figure}

\begin{figure}
\includegraphics[width=\textwidth]{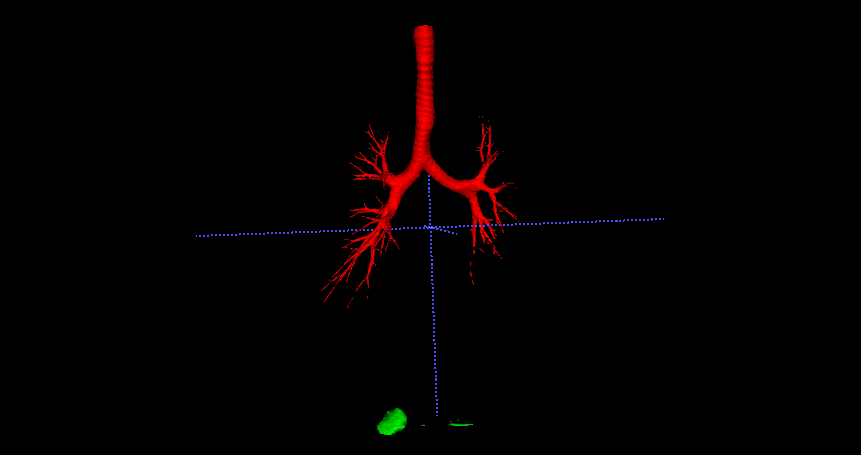}
\caption{Labeled airway tree segmentation for the validation case ATM\textunderscore292\textunderscore0000.}
\label{fig5}
\end{figure}

We implement a threshold of 0.5 on the output map $S_{fuse}$ to obtain the predicted mask for the 50 CT for validation. However, the predicted mask may contain spurious non-airway regions. Notably, the airway and non-airway regions can be generally topologically viewed in a Hausdorff space (see Fig. \ref{fig4} for example). We then refine the segmented results by removing the distinguishable large non-airway parts.  Due to the separability, we assign different labels to each separate region, for example, $1$ for the airway region and $2$ for the non-airway region as shown in Fig. \ref{fig5}. Similar post-processing methods can be found in \cite{ref_article6}.  Extracting the mask numerically labelled with $1$ and applying to the validation CT scans, we finally obtained the refined airway tree segmentation results. 

\section{Experiments and Results}
\begin{figure}
\includegraphics[width=\textwidth]{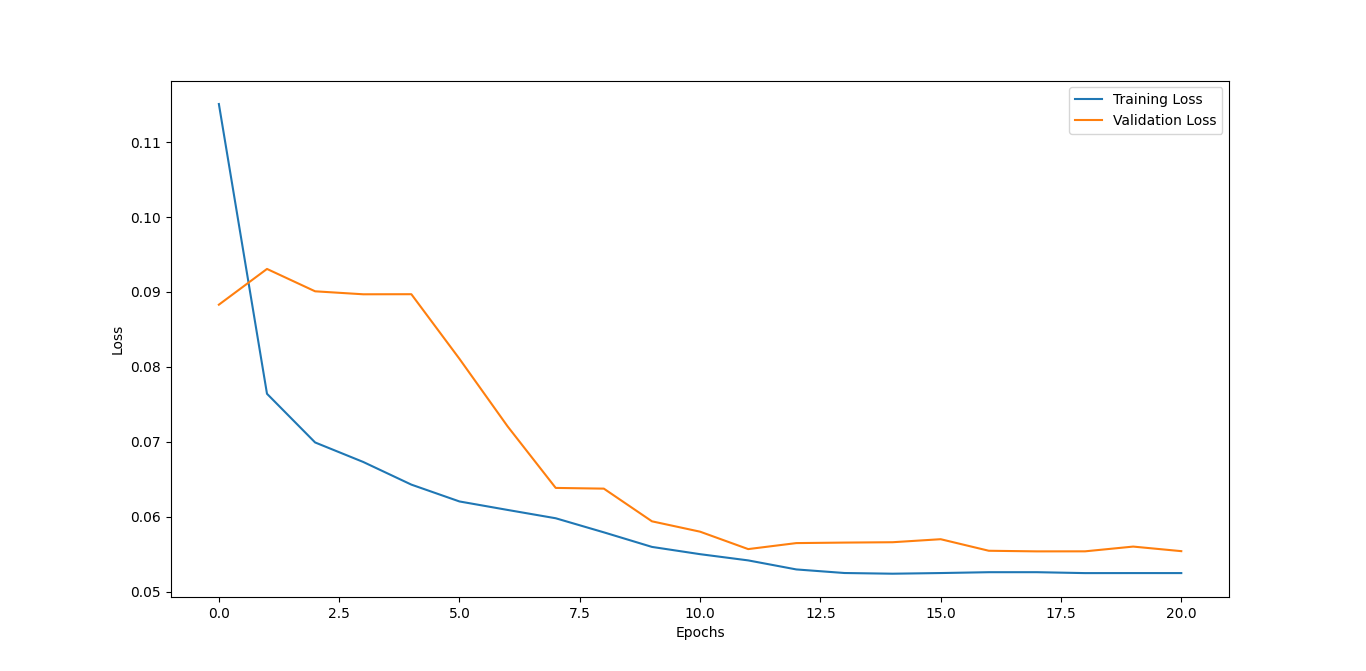}
\caption{Training loss and validation loss over epochs.}
\label{fig6}
\end{figure}

The U$^2$-net is implemented in the Pytorch framework and trained on NVIDIA GeForce GTX 1080TI with 11GB memory as well as RTX 3090 with 24GB memory.  The ADAM (Adaptive Moment Estimation) and SGD (Stochastic Gradient Descent) are chosen as the respective global and local optimizers. We employ the learning rate of $0.001$ with momentum of $0.9$ and set the batch size to 2. Also, we also choose $25\%$ of the input data of 299 CT scans as the test during the training of the network. 
We examine the training loss and validation loss over epochs. The training of one epoch takes about 10 hours. The training loss stabilizes at $0.0525$ while the validation loss converges to $0.0555$ both after around $12.5$ epochs as shown in Fig. \ref{fig6}.  We terminate the training of the network after 20 epochs and employ the thus-obtained parameters. 

%\begin{figure}[htpb]
%    \centering
%    \includegraphics[width=\textwidth]{loss_vs_epochs.eps}
%    \caption{Training loss and validation loss over epochs.}
%    \label{fig2}
%\end{figure}

  %Fig \ref{fig2} shows the loss versus epochs. The training loss converges to 0.0525 while the validation loss converges to 0.0555.  %The training stops until convergence of loss is reached. We then retrieve the model and the trained parameters with the minimum validation loss.

\begin{figure}[htbp!]
    \centering
    \includegraphics[width=\textwidth]{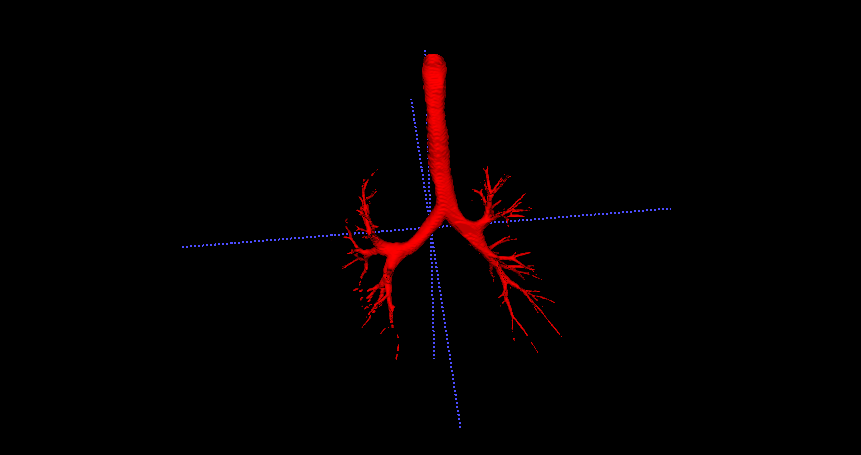}
    \caption{Final airway tree segmentation for the validation case ATM\textunderscore292\textunderscore0000.}
    \label{fig1}
\end{figure}

Fig. \ref{fig4} shows the unrefined  result of the predicted mask applying to the validation case ATM\_292\_0000. Following the post-processing method in Subsection \ref{sub2}, the refined airway segmentation of the case ATM\_292\_0000 is presented in Fig. \ref{fig1}.

The saliency function obtained from the U$^2$-net generally performs well on the validation CT scans. The observation shows that though the end part of the bronchial reflects certain weakness in connectivity, the main segmented airway tree including the tracheal and main sub-branches  behaves reasonably well in terms of connectivity and accuracy.

%Due to the regulation of ATM'22 grand challenge, we do not have access to the ground truth of the 50 validation CT scans.  Therefore, we cannot evaluate our trained model by calculating the four proposed evaluation metrics. The predicted masks of the 50 validation CT scans are submitted for evaluation by the challenge organizers.

\section{Conclusion}
We have obtained the airway tree segmentation of the given validation CT scans using the U$^2$-net. The experiment has indeed demonstrated that the U$^2$-net can generate a rather goodness of fit salicency probability function without any \emph{a priori} backbones requiring a reasonable GPU memory and computational cost. Our refined airway tree modeling shows good connectivity and accuracy by pure observation. Yet, we still need the ground truth to numerically examine the derived airway tree model.  Also, the post-processing treatment can also improve, for example, using the connectivity analysis given the ground truth of the validation data.  

%\subsubsection{Acknowledgements} Please place your acknowledgments at
%the end of the paper, preceded by an unnumbered run-in heading (i.e.
%3rd-level heading).


\begin{thebibliography}{8}
\bibitem{ref_article1}
Qin, X., Zhang, Z., Huang, C., Dehghan, M., Zaiane, O. R.,: U2-Net: Going deeper with nested U-structure for salient object detection. Pattern Recognition \textbf{106}, 107404 (2020)

\bibitem{ref_article2}
Lo, P., Van Ginneken, B., Reinhardt, J.M., et al.: Extraction of airways from CT (EXACT'09). IEEE Transactions on Medical Imaging \textbf{31}(11), 2093--2107 (2012)

\bibitem{ref_article3}
Zheng, H., Qin, Y., Gu, Y., et al.: Alleviating class-wise gradient imbalance for pulmonary airway segmentation. IEEE Transactions on Medical Imaging \textbf{40}(9), 2452--2462 (2021)

\bibitem{ref_article4}
Zhang, M., Yu, X., Zhang, H.,et al.: FDA: Feature Decomposition and Aggregation for Robust Airway Segmentation//Domain Adaption and Representation Transfer, and Affordable Healthcare and AI for Resource Diverse Global Health, pp. 25--34. Springer, Cham (2021)

\bibitem{ref_article5}
Garcia-Uceda, A., Selvan, R., Saghir, Z. et al.: Automatic airway segmentation from computed tomography using robust and efficient 3-D convolutional neural networks. Sci Rep \textbf{11}, 16001 (2021)

\bibitem{ref_article6}
Cheng, G., Wu, X., Xiang, W., et al.:Segmentation of the Airway Tree from Chest CT Using Tiny Atrous Concolutional Network. IEEE Access.\textbf{9}, 33583--33594 (2021)


\bibitem{ref_conf1}
Yu, W., Zheng, H., Zhang, M.,et al.: BREAK: Bronchi Reconstruction by gEodesic transformation And sKeleton embedding. In: 2022 IEEE 19th International Symposium on Biomedical Imaging (ISBI), pp. 1--5. IEEE (2022)

\bibitem{ref_conf2}
Qin, Y., Chen, M., Zheng H., et al.: Airwaynet: A voxel-connectivity aware approach for accurate airway segmentation using convolutional neural networks. In: International Conference on Medical Image Computing and Computer-Assisted Intervention, pp. 212--220. Springer, Cham (2019)

\bibitem{ref_conf3}
Milletari, F., Navab, N., Ahmadi, S.: V-net: Fully Convolutional Neural Networks for Volumetric Medical Image Segmentation. In: 2016 Fourth International Conference on 3D Vision (SDV), pp. 565--571. IEEE Computer Society, Los Alamitos (2016). \doi{10.1109/3DV.2016.79}

\bibitem{ref_conf4}
Jadon, S.: A survey of loss functions for semantic segmentation. In: 2020 IEEE Conference on Computational Intelligence in Bioinformatics and Computational Biology (CIBCB). IEEE, (2020)

\end{thebibliography}
%

\end{document}